\documentclass[apj]{emulateapj}
\usepackage{bm}
\newcommand{\bjdtdb}{${\rm {BJD_{TDB}}}$}

\newcommand{\mj}{${\,{\rm M}_{\rm J}}$}

\newcommand{\three}{3.6$\mu$m\ }
\newcommand{\four}{4.5$\mu$m\ }
\newcommand{\water}{$\mathrm{H}_2\mathrm{O}$\ }

\begin{document}

\title{The Broadband and Spectrally-Resolved H-band Eclipse of KELT-1b and the Role of Surface Gravity in Stratospheric Inversions in Hot Jupiters}

\author{Thomas G.\ Beatty\altaffilmark{1,2}, Nikku Madhusudhan\altaffilmark{3}, Richard Pogge\altaffilmark{4}, Sun Mi Chung\altaffilmark{4}, Allyson Bierlya\altaffilmark{5}, B.\ Scott Gaudi\altaffilmark{4,6}, \& David W. Latham\altaffilmark{5}}

\altaffiltext{1}{Department of Astronomy \& Astrophysics, The Pennsylvania State University, 525 Davey Lab, University Park, PA 16802, USA; tbeatty@psu.edu}
\altaffiltext{2}{Center for Exoplanets and Habitable Worlds, The Pennsylvania State University, 525 Davey Lab, University Park, PA 16802, USA}
\altaffiltext{3}{Institute of Astronomy, University of Cambridge, Madingley Road, Cambridge CB3 0HA, UK}
\altaffiltext{4}{Department of Astronomy, The Ohio State University, 140 W.\ 18th Ave., Columbus, OH 43210, USA}
\altaffiltext{5}{Harvard-Smithsonian Center for Astrophysics, Cambridge, MA 02138, USA}
\altaffiltext{6}{Jet Propulsion Laboratory, California Institute of Technology, 4800 Oak Grove Drive, Pasadena, CA 91109, USA}

\shorttitle{KELT-1b H-band Eclipse Observations}
\shortauthors{Beatty et al.}

\begin{abstract}
We present a high precision $H$-band emission spectrum of the transiting brown dwarf KELT-1b, which we spectrophotometrically observed during a single secondary eclipse using the LUCI1 multi-object spectrograph on the Large Binocular Telescope. Using a Gaussian-process regression model, we are able to clearly measure the broadband eclipse depth as $\Delta H=1418\pm94$\,ppm. We are also able to spectrally-resolve the $H$-band into five separate wavechannels and measure the eclipse spectrum of KELT-1b at $R\approx50$ with an average precision of $\pm135$\,ppm. We find that the day side has an average brightness temperature of $3250\pm50$\,K, with significant variation as a function of wavelength. Based on our observations, and previous measurements of KELT-1b's eclipse at other wavelengths, we find that KELT-1b's day side appears identical to an isolated 3200\,K brown dwarf, and our modeling of the atmospheric emission shows a monotonically decreasing temperature-pressure profile. This is in contrast to hot Jupiters with similar day side brightness temperatures near 3000\,K, all of which appear to be either isothermal or posses a stratospheric temperature inversion. We hypothesize that the lack of an inversion in KELT-1b is due to its high surface gravity, which we argue could be caused by the increased efficiency of cold-trap processes within its atmosphere.   
\end{abstract}

\section{Introduction}

The atmospheres of hot Jupiters are intricate multi-parameter systems. An exhaustive understanding of their atmospheres would include the effects of external irradiation, clouds, varying sedimentation efficiency, surface gravity, internal heat, the bulk metallicity, specific abundances, and rotation \citep[e.g.,][]{burrows2006,fortney2008,showman2009}. Needless to say, the complicated interplay between these cooperating and competing affects makes it difficult to isolate the individual parameters setting hot Jupiters' observed atmospheric properties.

As in stellar astronomy, we have two strong tools with which to understand hot Jupiters' complicated atmospheres: differential comparisons between planets to isolate and vary one atmospheric parameter of interest, and the detailed characterization of individual planets using their atmospheric spectra.

As one particular example of a differential comparison, consider the presence of a stratospheric temperature inversion in a hot Jupiter's atmosphere. Early theoretical predictions indicated that most, or all, hot Jupiters should have TiO- or VO-driven temperature inversions as long as their day sides were hotter than the 1900\,K condensation point of TiO/VO gas \citep{fortney2008}. While early broadband measurements of exoplanets' emission spectra seemed to bear this out \citep[e.g.,][]{knutson2008}, reanalyses of these early observations \citep{diamondlowe2014} and observations of other systems \citep[e.g.,][]{madhu2014,crossfield2015} have not yielded any significant detections of thermal inversions in hot Jupiters with day sides cooler than around 2500\,K.

Recently, however, \cite{haynes2015} reported the detection of an inversion in the emission spectrum of WASP-33b. The bulk properties of WASP-33b are presumably similar to other the hot Jupiters searched for inversions, except that the day side brightness temperature of WASP-33b is on average 3300\,K. Based on this, \cite{haynes2015} suggested that stratospheric temperature inversions may only be clearly present in hot Jupiters with very hot, $\sim$3000\,K, day sides. If correct, this indicates that there are additional aspects beyond just the condensation point of TiO/VO that govern the presence of inversion.

To make good differential comparisons, we also need to make good observations of the planets themselves. The atmospheric spectra of exoplanets are usually measured along the planetary terminator via transmission spectroscopy, or integrated over the day side via eclipse observations. Most ground-based exoplanet eclipse measurements observe in one of the broadband near infrared (NIR) filters. These broadband data have shown us the diversity of albedos and recirculation regimes that exist in hot Jupiters' atmospheres \citep[e.g.,][]{cowan2011}. However, the low-spectral resolution of these broadband measurements ($R\approx5$) has made it difficult to use them to draw strong conclusions about composition or the vertical temperature-pressure structure in individual atmospheres. In large part this is a reflection of the intrinsic complexity of hot Jupiters' atmospheres, and the difficulty in constraining multi-parameter models using relatively sparse datasets.

So far, almost all of spectrally-resolved ($R\approx50$) exoplanet eclipse observations have been made using space-based observatories. Based purely on photon-noise expectations, large ground-based telescopes should be able to easily exceed the performance of the space telescopes, but atmospheric affects have proven to be a large road-block in attempts to collect spectrally-resolved data from the ground. This has limited the clear detection of a ground-based eclipse spectrum to just two systems, HD 189733b \citep{swain2010,mandell2011,waldmann2012} and WASP-19b \citep{bean2013}, with another tentative detection of an eclipse spectrum from WASP-12b \citep{crossfield2012}.

After some early non-detections \citep{richardson2003}, the first clear measurement of an exoplanet's emission spectrum from the ground was accomplished by \cite{swain2010}, using SpeX on the NASA Infrared Telescope Facility (IRTF) to observe HD 189733b in $K$ and $L$ at $R\approx80$ in single-slit mode. Nevertheless, as an example of the difficulty of ground based eclipse spectroscopy, \cite{mandell2011} observed an $L$-band eclipse of HD 189733b using Keck/NIRSPEC at very high resolution ($R=27,000$) with a single slit to confirm \cite{swain2010}'s measurements, but were unable to reproduce their results. \cite{mandell2011} cautioned that telluric water emission in the $L$-band could be the cause of \cite{swain2010}'s signal. \cite{waldmann2012} later collected new SpeX observations of the system that showed an $L$-band eclipse spectrum stronger than expected from telluric features. Notably, while \cite{waldmann2012} observed using a single spectrosopic slit, they used one large enough to include a nearby star that they used as a simultaneous reference. This improved the quality of their results compared to \cite{swain2010}.

\cite{crossfield2012} also used a single slit in their measurements of WASP-12b's eclipse, which they also observed using SpeX. The lack of a reference star unfortunately limited the absolute precision of their measured eclipse depths, though they were able to make a tentative ($\sim1\sigma$) detection of WASP-12b's eclipse spectrum by differencing the individual spectral wavechannels against the broadband eclipse light curve. \cite{crossfield2012} also includes a detailed and illuminating discussion of their reduction and analysis techniques, which clearly elucidates the difficulties facing these types of observations from the ground.

The first ground-based eclipse observations using a multi-object spectrograph were conducted by \cite{bean2013} using the MMIRS instrument on one of the Magellan telescopes. By placing spectral slits over other nearby stars, \cite{bean2013} were able to use the other stars as differential comparisons, and thus remove much of the systematic noise in their light curve. Using two combined eclipse observations, this allowed them to measure the $H$ and $K$ emission spectrum from WASP-19b with a average precision of 380\,ppm at $R\approx15$. 
 
In an attempt to expand the number of exoplanets for which we have ground-based eclipse spectra, we spectrophotometrically observed an eclipse of the transiting brown dwarf KELT-1b \citep{siverd2012}. Of the twelve known transiting brown dwarfs, KELT-1b is the most irradiated and around the brightest host star, a combination which allows for high signal-to-noise ratio atmospheric characterization observations. KELT-1b is a 27\mj\ brown dwarf on a short, 1.217 day, orbit around a 6500\,K F5V host star. The system has a model-dependent age of $1.75\pm0.25$\,Gyr, which would make KELT-1b a T2 dwarf with an effective temperature of 1200K if it were an isolated field object. Previous observations of KELT-1b's eclipses in $z'$ \citep{beatty2014}, $Ks$ \citep{croll2014}, and at \three and \four \citep{beatty2014} have measured day side brightness temperatures considerably hotter than this, at approximately 3200\,K.

We were particularly interested in observing KELT-1b due to its extremely high surface gravity of 22 times that of Jupiter. We wished to make a differential comparison of KELT-1b's eclipse spectrum against other very hot Jupiters such as WASP-33b, to observationally investigate the role surface gravity plays in setting the atmospheric properties of hot Jupiters and irradiated brown dwarfs. Additionally, we were able to use the LUCI1 multi-object spectrograph on the Large Binocular Telescope for our observations. Similar to the \cite{bean2013} observations, this allowed us to use multiple spectral slits over multiple stars in the instrument's field of view. Using these stars as simultaneous spectrophotometric references allowed us to eliminate much of the systematic trends in our time series, and allowed us to approach the typical precision and spectral resolution of space-based observations in our spectrally-resolved data. 

\section{Observations and Data Reduction}

We observed a single $H$ secondary eclipse of KELT-1 ($H=9.534$) on the night of UT 2013 October 26 using the LUCI1 near-infrared (NIR) multi-object spectrograph on the Large Binocular Telescope (LBT). The observations began at UT 2013 October 26 0153 and ran without interruption until UT 0704. This end time was 45 minutes earlier then scheduled, because the telescope guider failed. The predicted eclipse time was UT 0448, and the predicted eclipse duration was 2.70 hours \citep{beatty2014}. Our observations thus covered the entire duration of the predicted eclipse, plus 1.57 hours before the predicted ingress and another 0.92 hours after the predicted egress. At the start of the observations KELT-1 had an airmass of $z=1.24$, crossed through the meridian at UT 0503 with an airmass of $z=1.007$ and ended the observations at $z=1.10$. Conditions were photometric throughout. 

\begin{figure}[t]
\vskip 0.05in
\centerline{
\epsscale{1.2}
\plotone{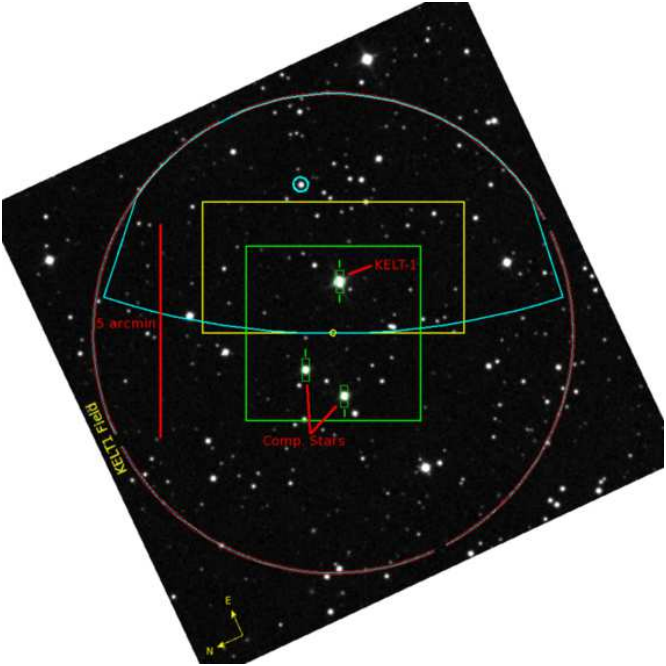}
}
\vskip -0.0in
\caption{The LUCI1 field of view, showing the positions of KELT-1, the two comparison stars, and their associated slits. We used large 10''$\times\,$30'' slits to avoid slit losses. Note the smaller narrow slits above and below the stars: these provided a telluric wavelength reference for each star's spectrum.}
\label{fov}
\end{figure}

For our observations we used a custom slit mask with LUCI1's N1.8 camera, the 210\_zJHK grating, an $H$ filter, and a central wavelength of $1.65\,\mu$m. A the time of our observations LUCI1 used a Hawaii-2 2048$\times\,$2048 pixel HgCdTe detector, which has since been replaced with an H2RG detector. The slit mask (Figure \ref{fov}) used three large (10''$\times\,$30'') slits to image KELT-1 and two nearby comparison stars, 2MASSJ00011914+3924347 (``Comp-1'') and 2MASSJ00011438+3924016 (``Comp-2''). These are the two brightest stars near KELT-1 that would simultaneously fall within LUCI1's 4'$\times\,$4' field-of-view. The two comparison stars have $H$ magnitudes fainter, but relatively close, to KELT-1's ($\Delta H=0.568$ and $\Delta H=0.817$, respectively) and similar NIR colors ($\Delta[J-K]=0.057$ and $\Delta[J-K]=0.006$, respectively). Comp-1 is located 2'.16 away from KELT-1, and Comp-2 is 2'.62 away. The orientation of the slit mask placed KELT-1 in the top half of our images, and the two comparison stars close together in the bottom half. In addition to the three slits for the stars, the mask had four smaller slits (2''$\times\,$10'') to observe sky lines and provide a wavelength reference.

We used an observing script that took groups of ten 60 second exposures. We read out using the \textsc{o2dcr} mode, which has a read time of four seconds. Including other overhead time, each full repeat of the script took approximately 13 minutes to complete. This length of time was driven by the recommendation of the LBT staff to refresh the guiding once every ten minutes. At the beginning of each script we therefore included a ``micro-offset'' of 0''.001 in the telescope guiding to trigger a reset of the guiding program. 

We calculated the $\mathrm{BJD}_{\mathrm{TDB}}$ time of each image from the $\mathrm{JD}_{\mathrm{UTC}}$ mid-exposure time given in each of the {\sc fits} image headers. Within each batch of ten images observed by a single run of our observing script, the images were evenly spaced every 68.5 seconds. Every tenth image the time between exposures increased to 85 seconds as the observing script and guider restarted. In total we acquired 267 images from JD 2456591.57875 to JD 2456591.79486.

In addition to our science images, we took five spectral flats the afternoon after the night of the eclipse. We used LUCI1's ``halo3'' flat lamp for these exposures, and the minimum exposure time of 4 seconds. All calibrations for the LUCI1 spectra were obtained using the onboard calibration system.  Flat field spectra were taken using the slit mask and the quartz-tungsten halogen (QTH) continuum lamp, and dark frames were acquired using the same exposure times as the individual target and flat-field images. We stacked multiple darks to create a median dark frame for the science and flat-field spectral images, and subtracted this median dark image from each frame. This removed most of the artifacts due to persistent hot pixels on the LUCI1 HgCdTe detector array.

Spectral flat fields required special handling because of the customs slit mask and the need to preserve the pixel-to-pixel noise characteristics as much as possible. The individual spectral flats were combined into a median `super flat'. For each slit (exoplanet host and comparison stars), we extracted the two-dimensional spectral flat trace and removed the wavelength-dependent color term due to a combination of the spectral response of the system and the spectrum of the QTH continuum calibration lamp. This produced a zero-color spectral flat that preserved the pixel-to-pixel gain variations within the spectral trace plus any residual non-color structure in the flat. This was done separately for each slit. The individual two-dimensional zero-color spectral flats for each slit were then merged back into a single master two-dimensional spectral flat field image.  

There is a small amount of residual flexure in the LUCI1 spectrograph that is not compensated for by the internal flexure compensation system. This is because of the very long visit time (5.2 hours), causing the spectrograph to swing through a wide range of orientations relative to gravity, which stresses the fidelity of the internal lookup tables used by the open-loop LUCI flexure compensation system. By comparing all of the science spectra from the entire run, we created a mask that isolated only those parts of the spectral traces that are common to all images.  This mask was applied to the merged zero-color spectral flat image, and used to perform the flat field correction for each of the science images. These final dark- and flat-corrected science images were used in the subsequent analysis of the secondary eclipse.

We did not attempt to correct cosmic ray hits in our images. This was motivated by a desire to remain as close to the actual observations as possible, and we accepted that images with cosmic rays in the spectral extraction region would appear as outliers. We did generate a bad-pixel mask from the median dark image, which we used to mask out the effected pixels from our spectrophotometric extraction process.

We used wide 10'' slits on KELT-1 and the comparison stars so as to avoid slit losses during the observations, but this had the additional affect of creating broad telluric features in our science images (e.g., the top panel of Figure \ref{data}). To perform precise photometry on all three stars, we needed to remove these broad telluric ``plateaus'' from our spectra without modifying the flux received from the stars.

\subsection{Telluric Background Subtraction}

To remove the telluric features from our data, we first determined the median sky background in regions away from the stellar light. We began by taking smaller strips of each science image around each of the stellar images, and worked with these individual sub-images. For each of our three stellar slits, the strip sub-images were 110 pixels tall, and the width of the detector along the dispersion axis. We centered the strips on their respective stellar traces from the first image of the night. Since the images of the three slits drifted by approximately 2 pixels on the detector over the course of our observations, this means that the fixed image strips include slightly different portions of the slit images as a function of time. As the telluric background appeared to be constant along the spatial direction of our slits, we did not consider the affect of this small slit movement on our results.

\begin{figure}[t]
\vskip 0.05in
\centerline{\epsscale{1.4}\plotone{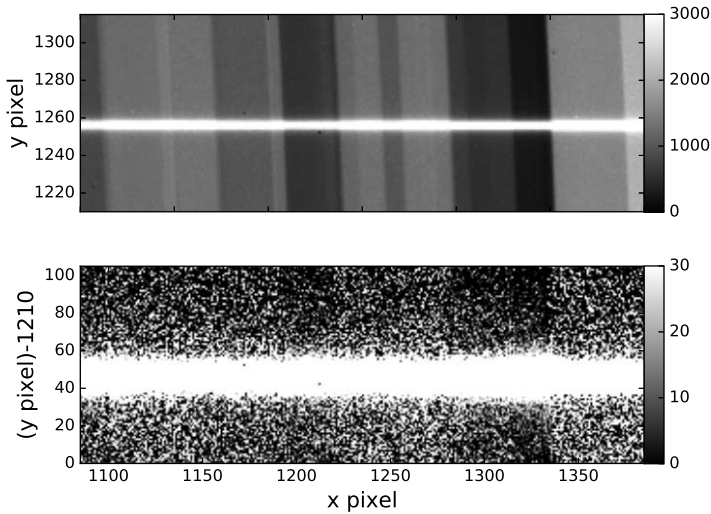}}
\vskip -0.0in
\caption{An example portion of a science image with (top) telluric background, and after subtraction (bottom). Note that the scale in the bottom panel is 100 times smaller than the top panel. The median subtracted background was about 1.5 DN/pixel, on average, compared to peak counts of approximately 8000 DN/pixel from the target stars.}
\label{data}
\end{figure}

We defined the background portions of each 110-pixel strip to be from row 0 to 20 and then from row 65 to 95. Due to the tilt of the telluric background in each strip (Figure \ref{data}), a straight median of these two background regions in each strip would smear out many of the features of interest. We therefore needed to rectify the background regions of each strip. To do so, we used the bottom row in each strip as a reference row, and linearly shifted each subsequent row in a strip's background regions such that it lined up with this reference row. We used a cubic interpolation to allow for sub-pixel shifts. The red crosses in the top panel of Figure \ref{data} demonstrate the shift values for the KELT-1 strip in the first image of the night. We then took the interpolated-and-shifted background rows and median combined them to determine the median telluric background in each strip.

We next subtracted the median background from the area of each strip around the stellar light. To do so, we constructed an appropriately tilted median background estimate near the stars by, first, fitting a quadratic polynomial to the shift values we determined in the background regions. We used these predicted shifts for each row near the star light to interpolate-and-shift our previously determined median background to match. This allowed us to construct a model of the tilted background, by extrapolating from both the level of background and its tilt, in regions of our 2D spectra away from the stellar light and use it subtract the background near the stars.

The bottom panel of Figure \ref{data} shows an example result of this subtraction process on a portion of the KELT-1 strip in the first image of the night. In the regions of the strip away from the star, the median of the subtracted background is 3.2 DN/pixel for this example image, and was on average 1.3 DN/pixel for the KELT-1 strips, 1.5 DN/pixel for Comp-1, and 1.9 DN/pixel for Comp-2. 

In addition to the interpolate-and-shift method that we have just described, we also experimented with the upsample-rectify-downsample telluric subtraction method described in \cite{stevenson2014}. We found the photometry from the two sets of differently reduced images to be qualitatively the same. Though there were small differences between individual measurements, the photometry from both reduction methods were consistent within $1\sigma$ of each other. We do note that \cite{stevenson2014}'s method completed in about one day, while our subtraction method took a little under a week to finish.

\subsection{Wavelength Calibration}

Our primary concern regarding the wavelength calibration of our spectra was to match wavelength ranges between KELT-1 and the two comparison stars during the spectrophotometric extraction, and to maintain this specific wavelength range over the course of the observations as the stars moved within their slits. To do so, we first used the telluric lines imaged by each of the narrow sky slits to determine an initial calibration, and then linearly offset this to match the positions of the stars, which were not centered in their wide slits, by identifying lines in the stellar spectra.

\begin{figure}[b]
\vskip 0.05in
\centerline{\epsscale{1.3}\plotone{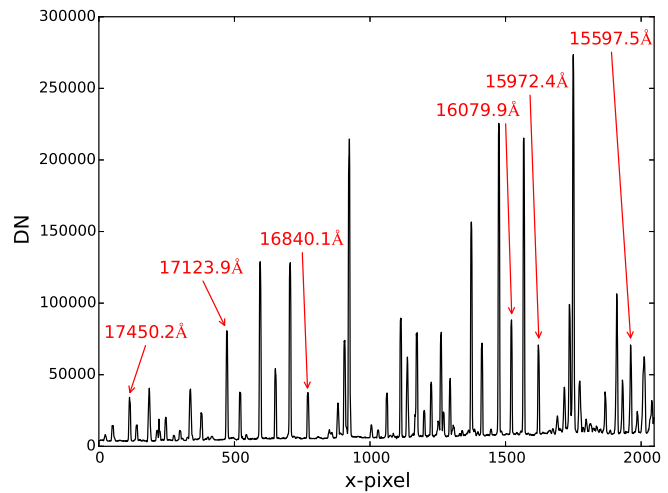}}
\vskip -0.0in
\caption{An example of the 1D telluric spectrum we used for our initial wavelength and dispersion calibration. Nearly all of the lines present are from atmospheric OH, and we have labeled the six lines we used in determining our 3rd-order polynomial fit for pixel as a function of wavelength for each image.}
\label{wcalib}
\end{figure}

To perform the telluric wavelength calibration, we used the four 2''$\times\,$10'' sky slits positioned immediately above and below the slits for KELT-1 and the two comparison stars. All four of these slits showed strong OH emission lines across their entire spectral range. As shown in Figure \ref{fov}, we had two slits above and below KELT-1, one slit above Comp-1, and one slit below Comp-2. In all of the sky slits the telluric lines were narrow plateaus approximately seven pixels wide. For each slit we extracted a rectangular region 20 pixels tall centered on the middle of the spatial axis of the sky slits, and that spanned the entire width of each image along the dispersion direction. We created 1D spectra of these extraction regions by simply summing along the detector columns. We did not attempt to rectify the telluric lines, as we judged the uncertainty introduced by their slope (a shift of approximately 0.2 pixels from the top of the extraction range to bottom) to be small relative to the flat-topped 7 pixel width of the lines themselves. 

For each of the four 1D sky spectra that came from each science image, we identified the same 6 isolated OH emission lines (Figure \ref{wcalib}) and tracked their position for each slit for all of our images. To determine the center of the lines along the dispersion direction we used the center-of-light centering method described by \cite{howell2006}. We fit the measured line centers by a third-order polynomial that gave x-pixel as a function of wavelength. The average standard deviation of these fits was 0.12 pixels, with variation from 0.2 pixels to 0.05 pixels over the course of the observations. These polynomial wavelength solutions allowed us to track the central wavelength and dispersion at the center of the stellar slits.

Due to small inaccuracies LBT's guiding, all three of our observed stars were neither in the center of their slits, nor did they remain stationary over the course of our observations. All three stars moved around in the dispersion and spatial axes, but for the purposes of wavelength calibration, we focus here on the movement of the stars along the dispersion direction.

\begin{figure}[t]
\vskip 0.05in
\centerline{\epsscale{1.4}\plotone{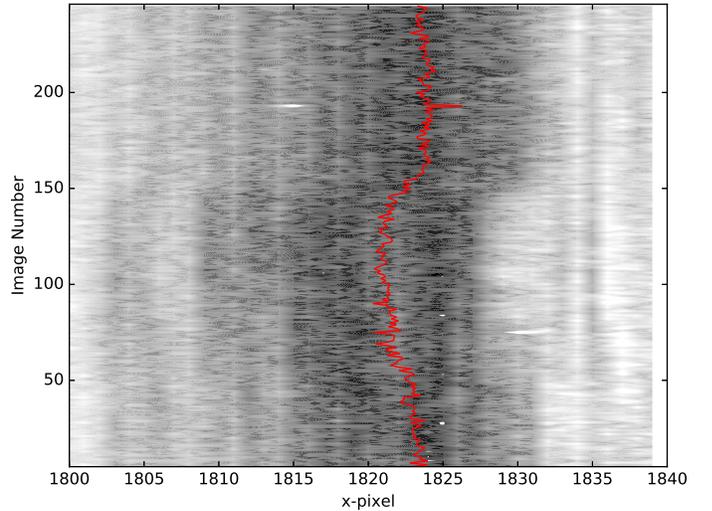}}
\vskip -0.0in
\caption{Motion of the Brackett 15-4 line in KELT-1's spectrum over image number. The red line is the fitted line centroid position. The sharp feature at about image no. 150 leads to a ``bump'' in our photometry halfway through the eclipse (Figure \ref{broadband}). We believe it was caused by a shift in the telescope or instrument optics shortly after crossing the meridian.}
\label{linetrack}
\end{figure}

The total movement along the dispersion axis is relatively small, no more than 3 pixels from the start locations, but trends from this movement were present in the initially uncorrected spectrophotometry. To translate the telluric wavelength calibration onto the sky slits, we treated the offset and motion of the stars along the dispersion direction of the detector as a linear offset to the telluric calibration; we did not attempt to correct for any variations in dispersion. To track the motion of the stars, we identified and centroided the Brackett 15-4 line in the spectra of KELT-1 and the two comparison stars. We created the 1D spectra we used for this line tracking by doing a simple spectral extraction in a rectangular area 30 pixels tall that was centered on the position of the stars in the first image of the night, and spanned the width of the detector. Figure \ref{linetrack} shows the resulting 1D spectra for KELT-1 stacked vertically and zoomed on the line we identified. The overplotted red line shows the line centroid locations, which we determined by fitting a Gaussian profile of variable width to the 1D spectra. The motion of the line over the course of the observations is clearly visible. Of particular note is the sudden movement of the line to the right around image number 150. We discuss this event in more detail in Section 3 and 3.5, as it introduced significant systematics into our photometry. For now, knowing the true wavelength of this line, we calculated an offset from the previously determined telluric wavelength solution, to transfer this solution to the true positions of the stars.

With the initial wavelength and dispersion calibration provided by the telluric emission imaged by the sky slits, and the offset provided by the stellar line tracking, we were able to define a constant wavelength region along the dispersion axis on the detector for each star in each image for the following spectrophotometric extraction. Recall from Figure \ref{fov} that to fit all three target stars within LUCI1's field of view, we were not able to align the stars along the dispersion direction. This means that the spectra for each covered slightly different wavelength ranges. Due to the motion of the stars on detector the exact range changed with each image, but was generally 1.55\,$\mu$m to 1.75\,$\mu$m for KELT-1, 1.52\,$\mu$m to 1.72\,$\mu$m for Comp-1, and 1.55\,$\mu$m to 1.75\,$\mu$m for Comp-2. We found that including Comp-1 in the detrending offered no improvement to the results, and so used only Comp-2. The total overlapping wavelength region available between KELT-1 and Comp-2 was therefore from 1.55\,$\mu$m to 1.75\,$\mu$m. This is slightly narrower than $H$-band, which is generally from 1.50\,$\mu$m to 1.80\,$\mu$m.

\section{Light Curve Extraction and Fitting}

We wished to measure both the broadband, $H$, eclipse depth of KELT-1b, and to spectrally resolve the eclipse to measure depth changes as a function of wavelength within the $H$-band. This led us to use slightly different spectrophotometric extraction techniques, and different fitting parameters, for the broadband and the spectrally-resolved light curves. Our fitting technique remained the same between the two types of measurements.

An initial examination of both the broadband and spectrally-resolved spectrophotometry showed strong correlated variability in our measured time series. First, we noticed a clear bump in the spectrophotometry around JD 2456591.70, or image number 150. As shown in Figure \ref{linetrack}, this corresponds to a sudden shift to the right in the position of KELT-1 within the spectral slit. This event also occurred at the same time that the entire spectral trace for KELT-1 deformed towards the right of our images, which we consider to be the root cause of the KELT-1's apparent motion within its slit. The change in the shape of the trace points towards an event within the optics of LUCI1, but we have no direct information about the optics during the observations. Our best theory, since the shift happened approximately 15 minutes after the telescope passed through the meridian, is that shortly after the meridian crossing some portion of LUCI1's optics shifted position slightly, causing the deformation of the trace, the shift of KELT-1's apparent position, and the bump in the spectrophotometry.

Second, we presume that the Hawaii-2 detector in use with LUCI1 at the time of our observations showed the same sort of linearity (and flux-dependent non-linearity) effects that are common to all NIR detectors. Unfortunately, no characterization of the linearity of the detector was done before it was replaced in the spring of 2015, and since the exact linearity corrections vary between NIR detectors, we did not consider it appropriate to use the linearity corrections derived for another chip. We therefore did not attempt any linearity correction to our science images. Since the peak counts in our images was about 8,000 \textsc{adu} pix$^{-1}$, which is about 12\% of the full-well depth in the LUCI detectors, presumably the actual linearity correction would have been relatively low.

Another possible source of systematic uncertainty in our observations is the possible stellar companion to KELT-1 discovered by \cite{siverd2012}. The companion is located 558 mas to the southeast of KELT-1, and has $\Delta H=5.90\pm0.10$ and $\Delta K'=5.59\pm0.12$. The companion is unresolved in our observations. Based on the the $\Delta H$ measurement of the companion to KELT-1, it contributes approximately 0.4\% of the system light in the wavelength range we observed. This is substantially below the fractional uncertainty we calculate for the broadband eclipse depth (about 7\%). We therefore chose to ignore its contribution to our observations. 

\subsection{Fitting via Non-parametric Gaussian Processes}

To deal with the correlated noise in our spectrophotometric time series, we choose to fit the eclipse data via a non-parametric Gaussian process (GP) regression. We initially attempted to fit the eclipse using a more traditional polynomial decorrelation against the major systematic noise variables (e.g., telescope focus and airmass). This polynomial detrending gave a fit to the eclipse that looked reasonable and had a formal uncertainty on the eclipse depth that was on the order of 5\%. Unfortunately,a closer investigation showed that the final measured properties of the eclipse changed significantly depending upon the choice of decorrelation function we made (i.e., linear, quadratic, etc.), and upon what specific variables we were decorrelating against (time vs. airmass, for example). The changes in our measurements of the eclipse properties were typically two to three times larger than the formal uncertainty we determined for any specific set of decorrelation choices. We therefore concluded that the correlated noise in our data was causing a measurement uncertainty that was not being correctly captured or dealt with by the polynomial decorrelation techniques. 

This conclusion prompted us to investigate using a GP to fit our eclipse data, since one way to conceptualize a GP is as a marginalization over many possible systematic noise models. More specifically, a GP models the observed data points as random draws from a multivariate Gaussian distribution about some mean function. This is in contrast, for example, to a $\chi^2$ fitting process, which models the data as random draws from a univariate Gaussian distribution. As a result, GPs are able to model the possible covariances between data points while remaining relatively agnostic about the precise functional form of the systematic noise model. 

GP methods were first used in the context of time series observations of exoplanets by \cite{gibson2012}, though they have a longer history in the general astronomical community \cite[e.g.,][]{way2006}. For a detailed mathematical introduction to GPs, we refer the reader to \cite{rasmussen2006}, and to Appendix A of \cite{gibson2012}.

We will define our GP model using the notation from \cite{gibson2012}. Let us describe our observed $N$ data points as a vector of observed fluxes, $f = (f_1,..., f_N)^T$, with a vector of corresponding observed times, $t = (t_1,..., t_N)^T$. Additionally, we will keep track of $K$ state parameters (e.g., airmass and telescope focus) at each time $t$ with the state vector $x = (x_{t,1},...,x_{t,K})^T$. For compactness, we will combine these state parameter vectors for each of our $N$ observations in the $N \times K$ matrix, $X$. Our GP model will use an eclipse model $E(t,\phi)$ as its mean function, where $\phi$ is the set of physical parameters describing the eclipse. Finally, let us define the multivariate Gaussian distribution underlying the GP model with the $N \times N \times K$ covariance matrix $\Sigma(X,\theta)$, where $\theta$ is the set of ``hyperparameters'' used to generate the covariance matrix from the $X$ state parameters. This allows us to write the joint probability distribution of our observed data $f$ as
\begin{equation}\label{eq:3110}
p(f|X,\theta,\phi)=\mathcal{N}[E(t,\phi),\Sigma(X,\theta)],
\end{equation}
where $\mathcal{N}$ represents the multivariate Gaussian distribution. Our GP model will therefore depend upon on an eclipse model $E(t,\phi)$, and a generating function -- referred to as the covariance kernel -- for the covariance matrix $\Sigma(X,\theta)$. We generated our GP covariance matrices and calculated the GP likelihoods using the George python package \citep{georgeref}.

While we used the same eclipse model to fit both the broadband and spectrally-resolved datasets, we needed to choose different covariance kernels for each set of observations. We describe our exact choice of covariance kernels in Sections 3.3 and 3.4.

\subsection{The Eclipse Model, Parameters, and Priors}

For the mean function of our GP model, we use a \cite{mandel2002} eclipse light curve. To generate our eclipse light curves we used the \textsc{batman} \citep{kreidberg2015} implementation of \cite{mandel2002} algorithm. We modeled the eclipse by assuming KELT-1b was a uniformly bright disk, with no limb-darkening, being occulted by the much larger KELT-1.

We parameterized the broadband eclipse model using the quantities
\begin{eqnarray}\label{eq:3210}
\phi_\mathrm{broad} &=& (T_{S,\mathrm{pred}},\log P,\sqrt{e}\cos\omega,\sqrt{e}\sin\omega, \\ \nonumber 
&& \cos i,R_P/R_*,\log a/R_*,\delta).
\end{eqnarray}
These eight parameters are the predicted time of the secondary eclipse ($T_{S,pred}$), the orbital period ($\log(P)$), $\sqrt{e}\cos\omega$, $\sqrt{e}\sin\omega$, the cosine of the orbital inclination ($\cos i$), the radius of the planet in stellar radii ($R_P/R_*$), the semi-major axis in units of the stellar radii ($\log[a/R_*]$), and the eclipse depth ($\delta$).

For seven of these parameters ($T_{S,\mathrm{pred}}$, $\log(P)$, $\sqrt{e}\cos\omega$, $\sqrt{e}\sin\omega$, $\cos i$, $R_P/R_*$, and $a/R_*$), we had strong prior expectations for their values from the KELT-1b discovery paper \citep{siverd2012} and the previously observed eclipses of the system \citep{beatty2014}. We did not have a prior expectation for the eclipse depth, $\delta$. 

We assumed Gaussian priors on the eclipse parameters for which we have previous information, which we note does not consider any possible covariances between the parameters. We used the average central values and $1\sigma$ uncertainties for $\log(P)$, $\sqrt{e}\cos\omega$, $\sqrt{e}\sin\omega$, $\cos i$, $R_P/R_*$, and $\log(a/R_*)$ from the two eclipses observed by \cite{beatty2014}, which were themselves based on the results of the free-eccentricity fit in \cite{siverd2012}.

For the predicted time of secondary eclipse, we took the straight average of the two eclipse times measured by \cite{beatty2014} to determine an epoch-zero eclipse time, and then used the period and associated period uncertainty of KELT-1b to predict the eclipse for the time of the our observations. 

Regarding the period of KELT-1b, observations of the transits of the system over the past several years by the KELT follow-up network have demonstrated to us that the period given in the discovery paper is slightly too long: the discovery ephemeris currently gives predicted transit times that are approximately 30 minutes late. To refine the ephemeris for our eclipse observations, we observed a full transit of KELT-1b using the 1.2\,m telescope at the Fred L. Whipple Observatory (FLWO) on the night of UT October 18 2013. 

We observed the transit using KeplerCam on the 1.2\,m telescope. KeplerCam has a single 4K$\times$4K Fairchild CCD with a pixel scale of 0''.366 pixel$^{-1}$, for a total FOV of 23'.1$\times$23'.1. We observed the complete transit of KELT-1b on the night of UT October 18 2013 in the SDSS $i'$ filter with 30 second exposures. We reduced the data using a light curve reduction pipeline outlined in \cite{carter2011}, which uses standard \textsc{idl} techniques. We fit the reduced light curve using \textsc{exofast} \citep{eastman2013} using priors on the transit parameters from \cite{siverd2012}.

\begin{figure*}[!t]
\vskip 0.05in
\centerline{\epsscale{1.2}\plotone{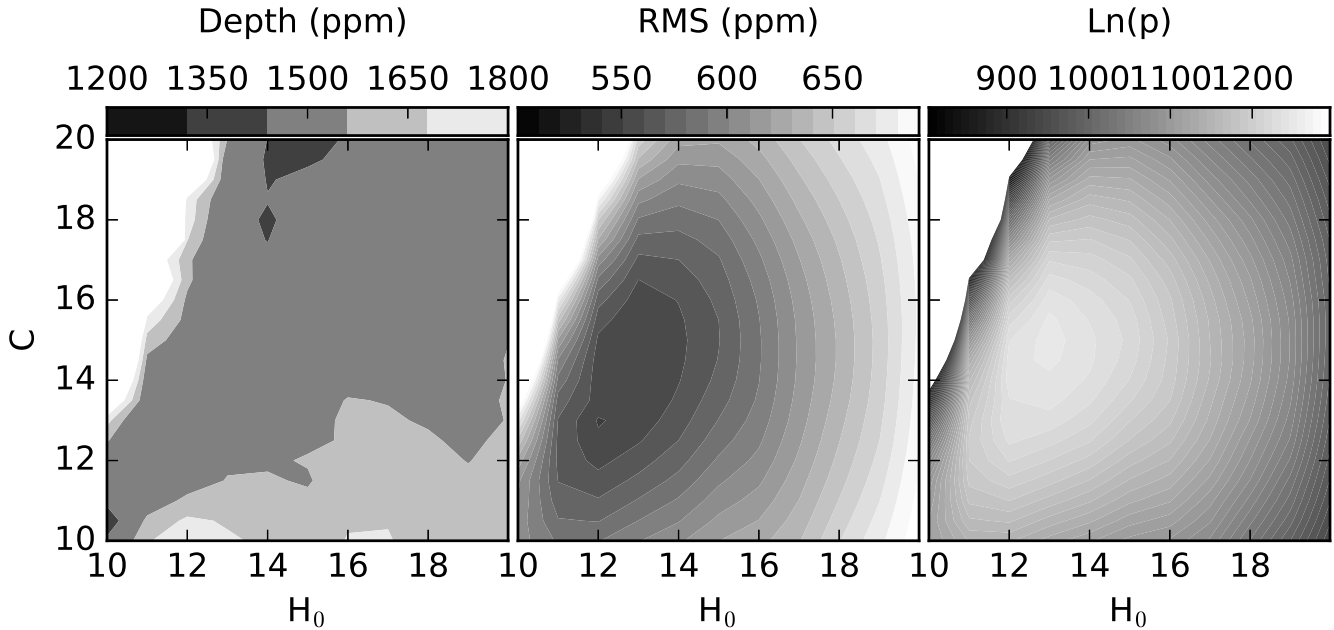}}
\vskip -0.0in
\caption{We scaled the spectrophotometric extraction aperture to the \textsc{m1z} value for each image, which roughly corresponds to the telescope's focus (Equation [\ref{eq:3310}]). $H_0=13$ and $C=15$ gave the combination of the lowest RMS and the highest $ln(p)$, and we use these values for our extraction. Note that in the far left panel the plotted contours correspond to multiples of our final $1\,\sigma$ uncertainty on the eclipse depth. The upper left area on all three panels gave extremely poor fitting results, and we excluded these from the plots.}
\label{apervary}
\end{figure*}

The measured transit center time from these observations of $\mathrm{BJD}_{\mathrm{UTC}}=2456583.78435\pm0.0006$ allowed us to considerably refine the orbital period of KELT-1b. Using the transit center time given in \cite{siverd2012}, we determined a new orbital period of $P=1.217494\pm0.000004$ days. This is approximately 1.6 seconds, and 1.2\,$\sigma$, shorter than the discovery's paper period of $P=1.217513\pm0.000015$ days.

As we have no prior expectation for the value of $\delta$, we do not include a probability penalty for this term, and therefore implicitly assume a uniform prior. We allowed for negative values of $\delta$ in both the broadband and spectrally-resolved fits.

For the spectrally-resolved eclipse observations, due to the relatively low signal-to-noise of the spectrally-resolved data we used an eclipse model with depth and secondary eclipse time as the free parameters:
\begin{equation}\label{eq:3220}
\phi_\mathrm{spec} = (\delta,T_S).
\end{equation}
We fixed the other parameters of the spectrally-resolved eclipse model $(\log P,\sqrt{e}\cos\omega,\sqrt{e}\sin\omega,\cos i,R_P/R_*,\\\log a/R_*)$ to the values which we determined from our broadband eclipse measurements. For $T_S$, the secondary eclipse time, we imposed a Gaussian prior from the time and uncertainty we measured in the broadband data.

\subsection{Broadband Spectrophotometric Extraction and GP Parameters}

Using our previously determined wavelength calibration for KELT-1 and the comparison star Comp-2, we defined a constant region in wavelength space where both stars overlapped, from 1.55\,$\mu$m to 1.75\,$\mu$m, on our telluric subtracted images to extract our spectrophotometry. To create our final summed spectrophotometry, we first summed along a limited range of the detectors' columns within this constant wavelength region. We defined the vertical limits of the extraction area relative to the stellar trace in each column of the detector. That is, for a given column $x$, the vertical extraction limits were from $T(x)-H(\mathrm{M1Z})$ to $T(x)+H(\mathrm{M1Z})$, where $T(x)$ is the vertical position of the stellar trace in column $x$, and $H(\mathrm{M1Z})$ is the vertical half-height of the extraction area. Note that the height of the extraction area depended upon the \textsc{m1z} value in the \textsc{fits} headers for our images. We explain this in more detail below.   

To determine $T(x)$ for each of our three stars, we fit a 9th-order polynomial to the spectral traces in each image. We determined the positions of the traces by fitting variable-width Gaussian profiles along the detector columns once every 20 columns. This gave us 101 $(x,y)$ positions, which we used to determine the polynomial fit to the trace.

In both the horizontal and vertical directions on the detector the borders of our extraction areas did not fall along integer pixel values, but partially included these edge pixels. We determined the flux contribution from these partially covered edge pixels via linear interpolation.

To determine the appropriate value for $H$, the height of the extraction area, we initially trialled several constant values of $H$, but found that the resulting broadband time series was strongly anti-correlated with the \textsc{m1z} \textsc{fits} header values. \textsc{m1z} roughly captures the value of the telescope's focus, and visual inspection of our telluric-subtracted images showed that the vertical width of all three stars' spectra decreased with increasing \textsc{m1z}, and vice versa.

To account for the slightly varying focus of the telescope and the slightly varying width of our spectra, we therefore choose to use a variable value of $H$ that depended upon the value of \textsc{m1z} for each image. Specifically, we set $H$ to be
\begin{equation}\label{eq:3310}
H(\mathrm{M1Z}) = H_0 + C\left(\frac{\mathrm{M1Z}}{\overline{\mathrm{M1Z}}}-1\right),
\end{equation}
where $H_0$ is some baseline value, $\overline{\mathrm{M1Z}}$ is the average value of \textsc{m1z} over all our observations, and $C$ is some scaling constant.

To determine appropriate values for both $H_0$ and $C$, we tested values of $H_0$ from 10 to 20 pixels, in integer steps, and values of $C$ from 10 to 20 in steps of 0.5. For each combination of values, we extracted and summed the spectrophotometry for all three stars, and then performed the initial maximization stage of our fitting procedure, which we describe in Section 3.5. This gave us an initial fit to the eclipse light curve, and we recorded the root-mean-squared (RMS) scatter of the residuals, the $ln(p)$ value of the fit, and the depth of the eclipse.

As shown in Figure \ref{apervary}, we found optimal values of $H_0=13$ and $C=15$, which makes the angular half-height of the extraction box roughly $3\farcs25$. This gave the combination of the lowest RMS and the highest $ln(p)$ (Equation [\ref{eq:3520}]). Note in the right-most panel of Figure \ref{apervary} that the broadband eclipse depths from these trial fits remained relatively constant near our choice for the optimum aperture: the fitted depths have a scatter of 90 ppm, while our final broadband eclipse depth has an uncertainty of 94 ppm. We therefore do not consider that our final results are affected above our errors by the precise choice of $H_0$ and $C$. 

\subsubsection{Broadband Covariance Kernel and Hyperparameter Priors}

Recall that we are defining the multivariate Guassian distribution underlying our GP model by its covariance matrix, $\Sigma(X,\theta)$, and the covariance matrix's associated generating covariance kernel.  

For our broadband spectrophotometry, we adopted a relatively simple ``squared-exponential'' kernel using time as the sole input state parameter. This allowed us to generate a square $N \times N$ covariance matrix for each point-wise combination of observation times $t_i$ and $t_j$ as:
\begin{equation}\label{eq:33110}
\Sigma(t,\theta) = A_t\, \exp\left[-\left(\frac{t_i-t_j}{L_t}\right)^2\right],
\end{equation}
where $\theta=\{A_t, L_t\}$ are the hyperparameters setting the amplitude and the length-scale of the point-wise covariance.

A squared-exponential kernel is generally used in GP modeling to account for variations in the data that occur smoothly as a function of the input state parameter. This makes it a good general choice for a covariance kernel, as a squared-exponential is usually regarded as the simplest choice for a kernel.

\begin{figure*}[t]
\vskip 0.00in
\epsscale{1.1}
\plotone{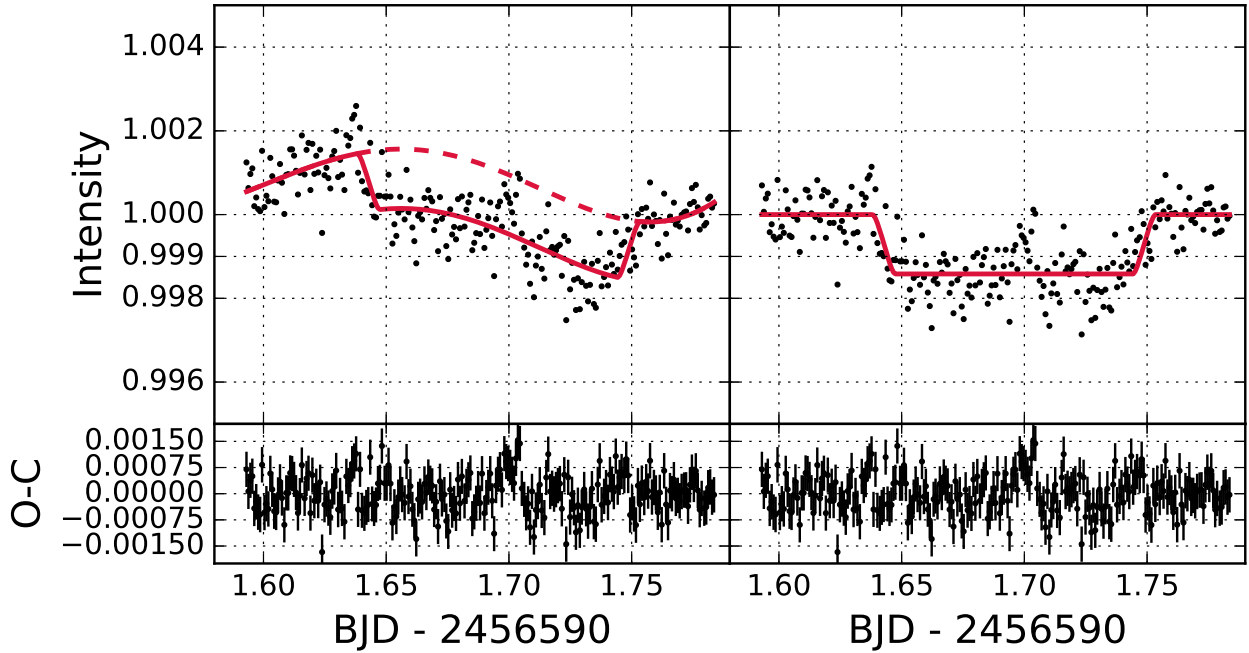}
\vskip -0.0in
\caption{Raw (left) and detrended (right) broadband photometry. The solid red line in the left panel shows the predictive mean of our Gaussian Process regression, which is a combination of the eclipse model and the noise model. The dashed red line in the left panel shows only the noise model. In the right panel, the red line is just the fitted eclipse model. The lower panels both show the residuals to our best fit model.}
\label{broadband}
\end{figure*}

In our fitting, we do not impose any prior on the kernel amplitude or the length-scale, other than requiring that the length-scale cannot be less than or equal to the predicted eclipse ingress and egress durations of 0.09 days. This lower limit on the allowable length-scale values ensures that the GP model does not treat the eclipse itself as correlated noise. 

\subsection{Spectrally-resolved Spectrophotometric Extraction and GP Parameters}

We extracted the spectrally-resolved spectrophotometry for KELT-1 and the comparison stars in the same way as we did for the broadband eclipse measurement, except that we divided the spectrum of each star into five evenly-spaced wavelength bins using our wavelength calibration. We used the same scaling relation with \textsc{m1z} to vary the extraction width as we did for the broadband extraction, and we set $H_0=13$ and $C=15$: the optimum values we found using the broadband results.

As with our broadband photometry, we detrended each individual wavechannel against the corresponding sections of the comparison stars' spectrophotometry. Again, we found that including the star Comp-1 in the detrending ensemble provide no improvement in the resulting light curves, and so we only used Comp-2 for detrending. This allowed the data to cover from 1.55\,$\mu$m to 1.75\,$\mu$m (i.e., Section 2.2).

\subsubsection{Spectrally-resolved Covariance Kernel and Hyperparameter Priors}

Since our spectrally-resolved light curves showed noticeably more correlated noise than the broadband spectrophotometry, we used a higher dimensional covariance matrix to model these additional noise sources. Visual examination of the spectrally-resolved data showed correlations between the measured flux and airmass ($X$). Again, in an attempt to keep our GP model as simple as possible, we used squared-exponential kernels to describe all of these covariances.

First, we included the same temporal covariance as in the broadband GP model with
\begin{equation}\label{eq:34110}
\Sigma_t(t,\theta) = \exp\left[-\left(\frac{t_i-t_j}{L_t}\right)^2\right].
\end{equation}
Next, we modeled the affect of changing airmass with the kernel
\begin{equation}\label{eq:34120}
\Sigma_\mathrm{air}(X,\theta) = \exp\left[-\left(\frac{X_i-X_j}{L_\mathrm{air}}\right)^2\right].
\end{equation}

We multiplicatively combined these two individual kernels together into one total kernel that we used for our spectrally-resolved GP model, 
\begin{equation}\label{eq:34150}
\Sigma_\mathrm{total}(t,X,\theta) = A_\mathrm{total}\,\Sigma_t\, \Sigma_\mathrm{air},
\end{equation}
where $\theta=\{ A_\mathrm{total}, L_t, L_\mathrm{air}\}$ are the four hyperparameters setting the properties of the covariance kernel. 

We combined the kernels via multiplication to effect the equivalent of a Boolean ``AND'' operation in the generation of our GP covariance kernel, in contrast to combining the kernels via addition, the method used in \cite{gibson2012}, which gives the equivalent of a Boolean ``OR.''  

During the fitting of the spectrally-resolved light curves we did not impose a prior on any of the amplitude terms, and only imposed hard lower limits on all of the coavariance length scales, to prevent the GP model from fitting out the eclipse, or attempting to fit the white noise. Specifically, we required $L_t > 0.09$ and $L_\mathrm{air} > 0.01$.

\subsection{Fitting Process and Results}

For both the broadband and spectrally-resolved data we used the same general fitting procedure to find the model parameters with the maximum likelihood and to estimate our parameter uncertainties. Specifically, we optimized the combined likelihood of our GP model and our priors. The natural logarithm of our GP model likelihood was
\begin{equation}\label{eq:3510}
\ln p_\mathrm{GP}(r|X,\theta,\phi) = -\frac{1}{2}\bm{r}^T\Sigma^{-1}\bm{r} - \frac{1}{2}\ln |\Sigma| - \frac{N}{2}\ln(2\pi),
\end{equation}
where $\bm{r}=f - \mathrm{E}(t,\phi)$ is the vector of the residuals of our observed data ($f$) from our eclipse model (E) defined in Section 3.2. This log-likelihood follows directly from our definition of the GP model in Equation (\ref{eq:3110}). 

Our combined log-likelihhod function combined the GP log-likelihood with our priors, as
\begin{equation}\label{eq:3520}
\ln p_\mathrm{tot}(\phi,\theta|f,X) = \ln p_\mathrm{GP}(r|X,\theta,\phi) + \sum \ln p_\mathrm{prior}. 
\end{equation}
We assigned Gaussian priors to all of the eclipse parameters, except the eclipse depth, as described at the end of Section 3.2 and Table \ref{tab:broadbandpriors}. Recall that we fit for all of the eclipse parameters with the broadband data, and only the eclipse time and depth for the spectrally-resolved data. For the hyperparameters, we imposed no priors, other than to impose a lower limit on the covariance length scales as described in Sections 3.3.1 and 3.4.1.

\begin{figure}[t]
\vskip 0.00in
\centerline{
\epsscale{1.2}
\plotone{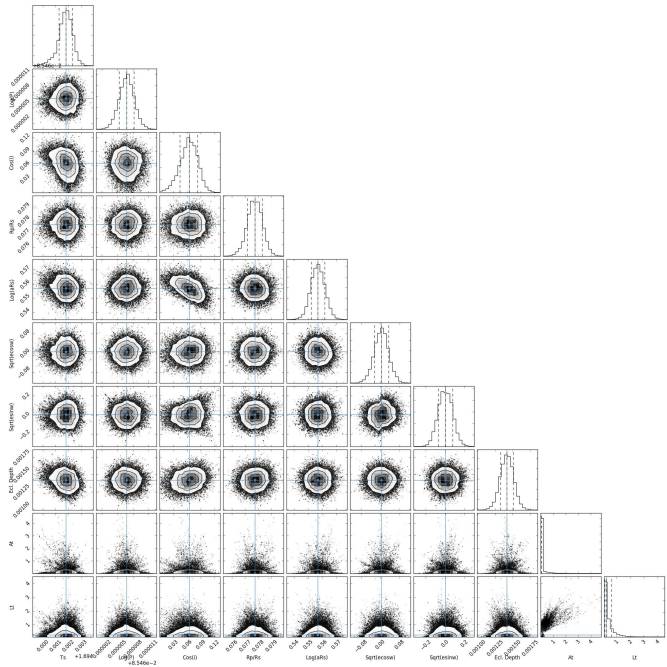}
}
\vskip -0.0in
\caption{Covariance plot for the broadband fitting process. We find that there is little covariance between our model parameters.}
\label{broadbandcovar}
\end{figure}

Our fitting process began by using a Nelder-Mead maximization to find an initial log-likelihood maximum to use as an estimate of the best fit. We then used MCMC to explore the parameter space around the log-likelihood maximum to determine uncertainties and to verify that we had correctly identified the global likelihood maximum. To perform the MCMC computations we used the \textsc{emcee} Python package. Our MCMC runs began with a 500 step burn-in, followed by 3000 production steps. At the end of the production run we verified that the MCMC chains had converged by calculating the Gelamn-Rubin statistic for each free parameters, and we judged the chains to have successfully converged if all the Gelman-Rubin statistics were smaller than 1.1. 

\begin{figure}[t]
\vskip -0.10in
\centerline{
\epsscale{1.2}
\plotone{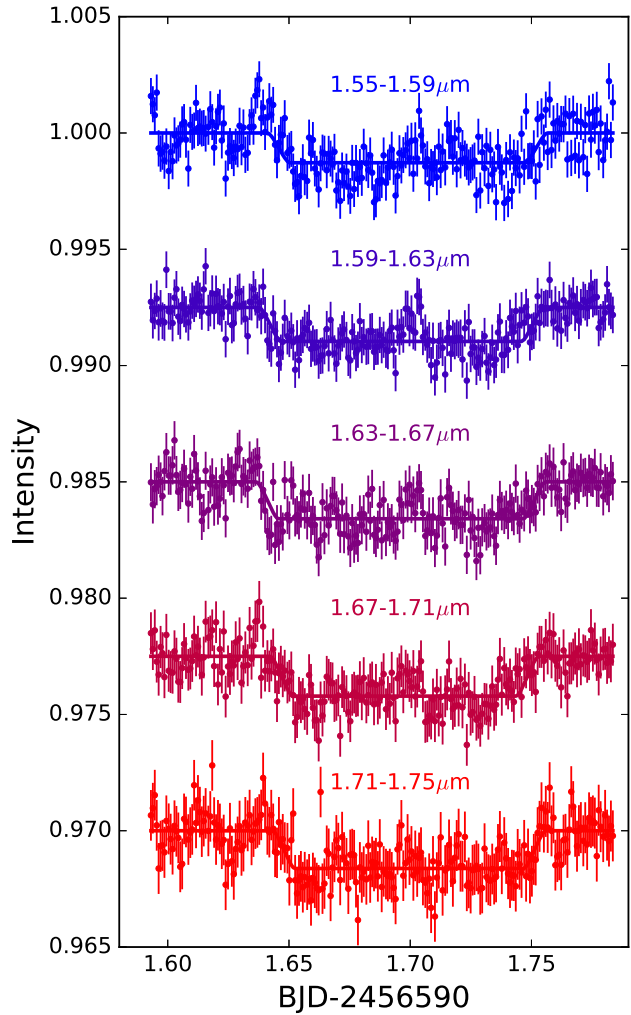}
}
\vskip -0.0in
\caption{The detrended sepctrally-resolved eclipse light curves. The flux in each individual wavechannel is plotted with an increasing offset from unity, for visual clarity.}
\label{specdepths}
\end{figure}

The results of our fit to the broadband data are shown in Figures \ref{broadband}, \ref{broadbandcovar}, and listed in Table \ref{tab:broadbandresults}. Our best fit gave $\delta=1418\pm94$\,ppm, This depth uncertainty is about four times what one would expect from photon noise noise alone. An Anderson-Darling test of the residuals gave a test statistic of 0.177, indicating that the distribution of the residuals was consistent with a Gaussian distribution. One feature to note in Figure \ref{broadband} is the ``bump'' that occurs around mid-eclipse. The timing of this event corresponded exactly with the jump in KELT-1's measured position shown about half-way up Figure \ref{linetrack}. This nicely demonstrates the necessity of keeping the target stars stationary -- or at least moving slowly -- on the detector during these types of observations.

\begin{figure*}[t]
\vskip 0.00in
\centerline{\epsscale{1.2}\plotone{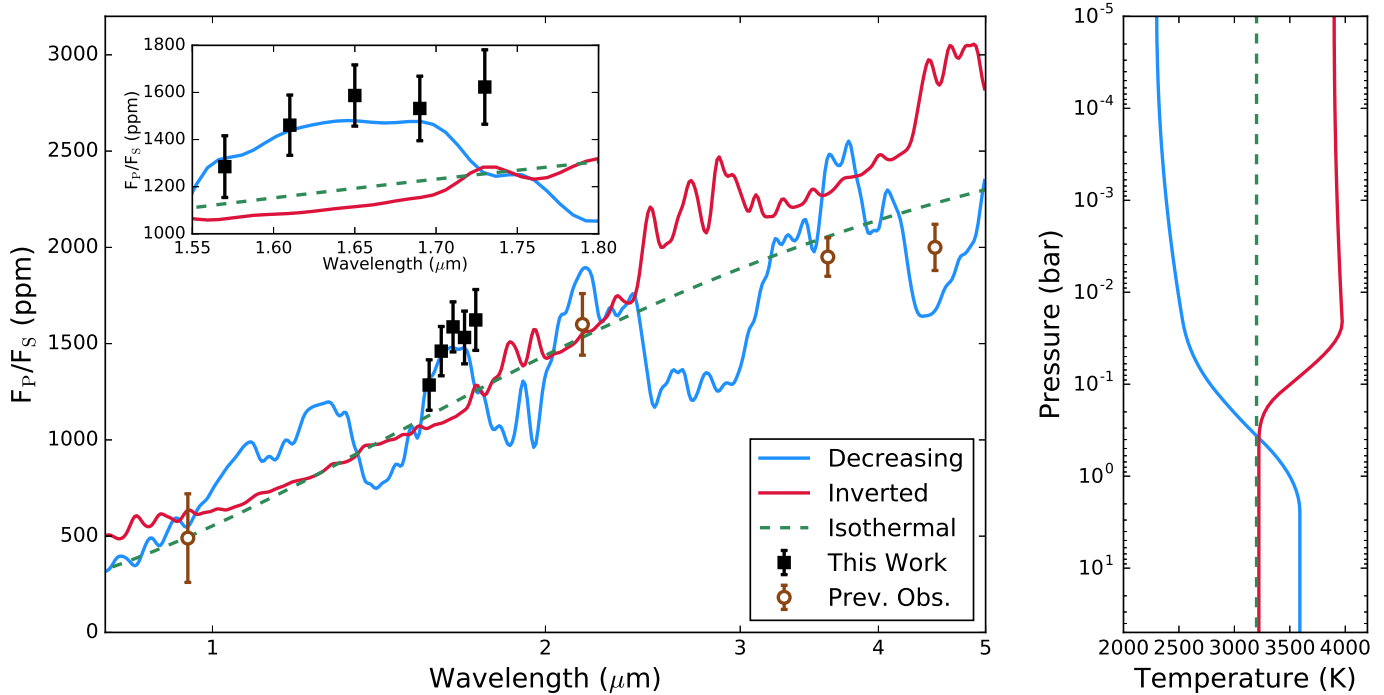}}
\vskip -0.0in
\caption{Our modeling of KELT-1b's $H$ emission spectrum and previous the broadband eclipses measured by \cite{beatty2014} and \cite{croll2014} using \cite{madhu2009}'s planetary atmosphere models favors a monotonically decreasing temperature-pressure (TP) profile. We exclude an 3250\,K isothermal TP profile at $3.3\sigma$ and an inverted TP profile at $8\sigma$.}
\label{modelatmo}
\end{figure*}

Figures \ref{specdepths} and Table \ref{tab:specresults} show the results of our fits to the spectrally-resolved eclipses. We clearly detected the eclipse in each wavechannel, but did not see a shift in the eclipse timing above the measurement uncertainties. We computed an Anderson-Darling test statistic for the residuals in each wave channel of the spectrally-resolved eclipse measurements, and found that all five were consistent with a Gaussian distribution. The uncertainties in the depth measurements were four times what one would expect from photon noise alone.

The average brightness temperatures of the spectrally-resolved eclipses corresponds to $3420\pm70$\,K, which is $2.3\,\sigma$ higher than the average day side brightness temperature of $3220\pm50$\,K one calculates using all of the available eclipse measurements.

\section{Modeling and Discussion}

KELT-1b is a relatively unique object, in that it is one of the twelve known brown dwarfs in a close orbit around a single main sequence star. This makes the external atmospheric forcing from the incoming stellar irradiation of KELT-1 similar to hot Jupiters, while the mass of KELT-1b and its implied internal energy are similar to field brown dwarfs. We therefore approached the analysis of KELT-1b's measured eclipse depths from two different directions.

\subsection{Atmospheric Modeling Results}

First, we modeled our spectrally-resolved eclipse depths, together with the previous eclipse measurements at other wavelengths, using the hot Jupiter atmospheric model described in \cite{madhu2009} and \cite{madhu2012}. This model is comprised of a 1D plane parallel atmosphere in hydrostatic equilibrium and local thermodynamic equilibrium (LTE). The emergent spectrum is computed using a 1D line-by-line radiative transfer solver in the planetary atmosphere. The atmospheric temperature-pressure (TP) profile and chemical composition are free parameters of the model, with 6 parameters for the TP profile and a parameter each for each chemical species included in the model. The range of molecules considered and the sources of opacity are discussed in \cite{madhu2012}. 

\begin{figure*}[t]
\vskip 0.00in
\centerline{\epsscale{1.2}\plotone{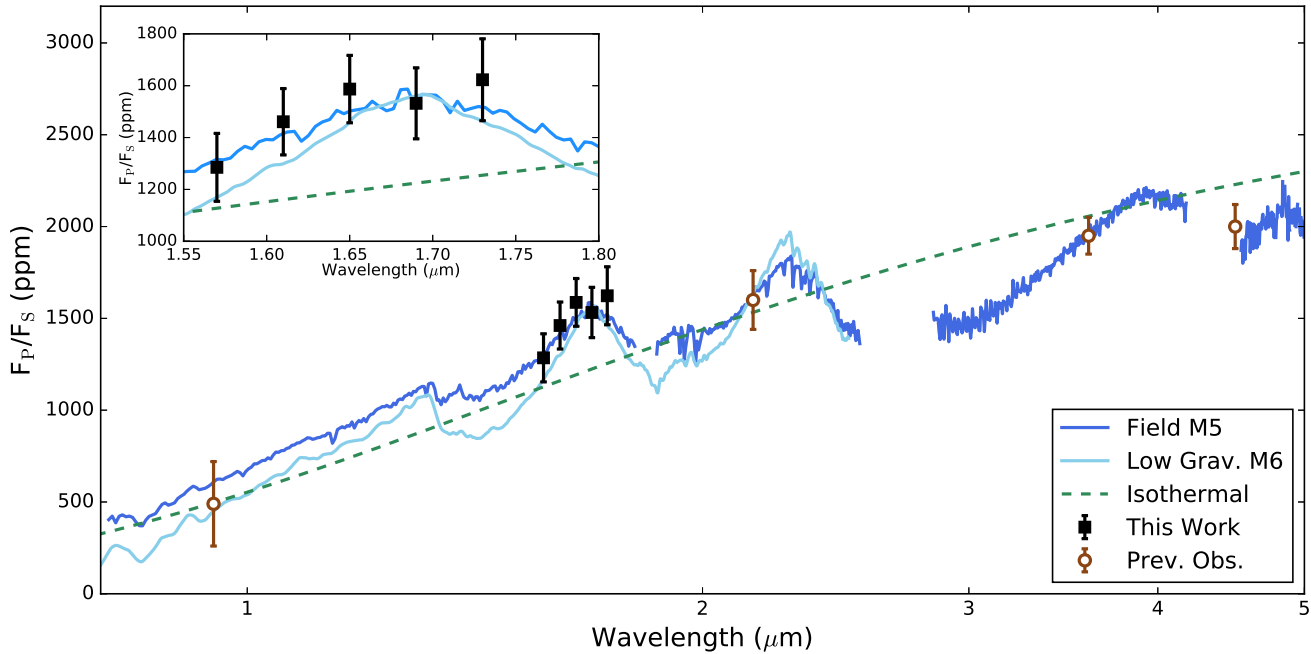}}
\vskip -0.0in
\caption{Measured thermal emission from KELT-1b from this paper, \cite{beatty2014}, and \cite{croll2014}. Also plotted are the best fit 3250\,K blackbody curve and template spectra for a high gravity field brown dwarf and low gravity brown dwarf. The normalization of the template spectra are computed using model predictions for the $H$ brightness of these objects, and is not a free parameter. The striking agreement between the observed emission spectrum of KELT-1b's day side and the field templates is another indication that the TP profile in KELT-1b's upper atmosphere is monotonically decreasing.}
\label{emission}
\end{figure*}

In the present work, our goal was to obtain a general understanding of the model parameter space consistent with the data rather than finding a unique best fit. We therefore assumed three different cases for the TP profile, and then fit for the best models that generally represented these three cases. The three TP profiles we considered were one that was monotonically decreasing, a profile with a thermal inversion, and an isothermal TP profile. We fixed the composition to be that predicted for chemical equilibrium with Solar elemental abundances for all three TP profiles, since we found that letting the abundances vary did not affect our conclusions regarding the temperature profile of KELT-1b's atmosphere.

Of the three TP profiles we considered, our modeling preferred a monotonically decreasing TP profile in KELT-1b's upper atmosphere using Solar values for the bulk atmospheric metallicity (Figure \ref{modelatmo}), with a best fit of $\chi^2=12.97$ for eight degrees of freedom ($\chi^2/\mathrm{dof}=1.62$). We exclude an 3250\,K isothermal TP profile at $3.9\sigma$ ($\chi^2/\mathrm{dof}=3.94$) and an inverted TP profile at $8.2\sigma$ ($\chi^2/\mathrm{dof}=11.50$) at Solar metallicities.

Visually, the lack of a thermal inversion is evident even without considering the model spectra. The $H$-band typically probes the continuum in a planetary spectrum due to lack of strong molecular absorption features in this spectral range \citep{madhu2012}. As such, the brightness temperature in the $H$-band is representative of the deepest layers of the observable atmosphere, below which a combination of several factors, such as increased density and collision-induced absorption, makes the atmosphere inaccessible in the visible and infrared. On the other hand, the Spitzer IRAC bands at \three and \four contain strong molecular bands due to all the prominent molecules expected in H$_2$-rich atmospheres.\footnote{$\mathrm{H}_2\mathrm{O}$, CO, $\mathrm{CH}_4$, and $\mathrm{CO}_2$ depending on the elemental abundances and temperature.} In particular, at the approximately 3200\,K temperatures of KELT-1b's day side, \water and CO provide strong absorption in the \three and \four bands, respectively. When the atmospheric TP profile is monotonically decreasing, absorption absorption from these two molecules leads to lower brightness temperatures in these two bands than that observed in the $H$-band. This is indeed what the observations show. Conversely, if the TP profile were inverted with the temperature increasing outward, or if it were isothermal, the brightness temperatures at \three and \four would be greater than or equal to that in the $H$-band. Both these scenarios are ruled out by the data.  

\subsection{Brown Dwarf Spectral Typing and Surface Gravity Indicators}

Second, we compared the day side atmosphere of KELT-1b to high-gravity field brown dwarfs, and low-gravity young brown dwarfs, by comparing the measured eclipse depths to spectral templates. To do so, we used field object templates from the SpeX Library covering spectral types from M4 to T9 and young, low-gravity, templates from \cite{allers2013} covering spectral types from M5 to T9. Using a 6500\,K blackbody for the emission from the star KELT-1, we transformed these template spectra from normalized flux vs. wavelength to expected eclipse depth vs. wavelength. We normalized the resulting eclipse spectra using the absolute $H$ magnitudes for these objects predicted by the BT-Settl models \citep{allard2011} using the \cite{caffau2011} values for Solar abundances. There were therefore no free parameters in our spectral typing process: the relative luminosities of KELT-1 and the field spectra were set by the BT-Settl predictions. We used the observed $z'$ \citep{beatty2014} and $Ks$ \citep{croll2014} eclipse depths in addition to our spectrally-resolved depths to evaluate the fit to the spectral templates. 

We find that the day side of KELT-1b appears to have a field spectral type of M5$^{+1}_{-2}$, and a low-gravity spectral type of M6$^{+1}_{-3}$ (Figure \ref{emission}). This is almost entirely due to the spectral slope of the eclipse depths from 1.55\,$\mu$m to 1.65\,$\mu$m, which drives the template fitting towards relatively broad wings for the \water absorption feature at 1.4\,$\mu$m. Both the field and low-gravity templates fit the measured eclipse depths extremely well, with $\chi^2=1.34$ ($\chi^2/\mathrm{dof}=0.19$) for the field template and $\chi^2=4.39$ ($\chi^2/\mathrm{dof}=0.63$) for the low-gravity template, both with seven degrees of freedom.

The goodness of fits for both spectral templates is quite striking considering that we allowed for no free parameters in the comparison process. Both spectral types correspond to effective temperatures of approximately 3200\,K, which approximately the same as our average measured brightness temperatures for KELT-1b at these wavelengths. More generally, the exquisite agreement between the observed emission spectrum of KELT-1b's day side and the field templates is also indicative that the TP profile in KELT-1b's upper atmosphere is monotonically decreasing, since both the field and low-gravity brown dwarfs are expected to steady cool going towards lower pressures in the outer portions of their atmospheres. 

While our spectrally-resolved results appear to show the ``peaky H-band'' feature typically used to identify low-gravity objects at later spectral types, by M5 this feature and its associated H-cont spectral index have largely lost their power to distinguish between field and low-gravity brown dwarfs \citep{allers2013}. As a result, as illustrated by the inset in Figure \ref{emission}, our spectrally-resolved results are not sufficient to give a clear preference to a field or a low-gravity spectral template. 

\begin{figure}[t]
\vskip 0.00in
\centerline{\epsscale{1.4}\plotone{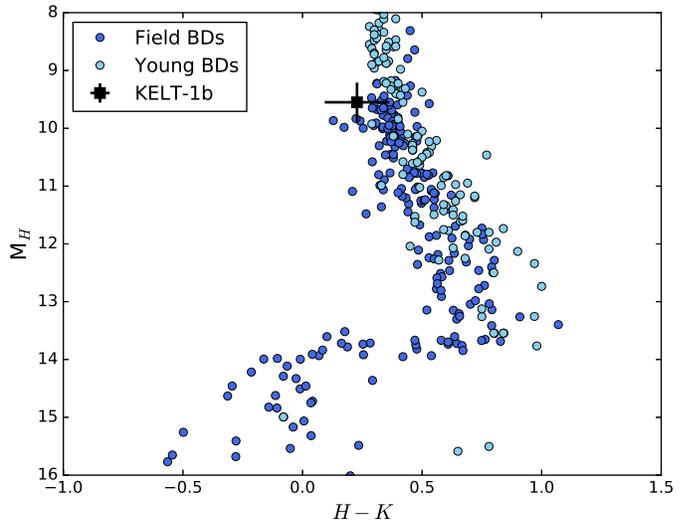}}
\vskip -0.0in
\caption{KELT-1b's day side has an $H-K$ color and and $H$-band brightness consistent with the brown dwarf color sequence. While KELT-1b's day side appears more similar to higher gravity field objects than the low gravity sequence, this is at a low ($1.6\,\sigma$) significance. The field brown dwarf magnitudes are from \cite{dupuy2012} and the young brown dwarfs magnitudes are from \cite{gagne2015}.}
\label{cmd}
\end{figure}

The broadband $H-K$ colors of KELT-1b's day side are similarly non-discriminatory. Using the measured $H$ and $Ks$ eclipse depths for KELT-1b from this work and from \cite{croll2014}, and the distance for KELT-1 determined by \cite{siverd2012}, we compared the day side NIR colors of KELT-1b's atmosphere to the field \citep{dupuy2012} and young \citep{gagne2015} brown dwarf color-magnitude sequences (Figure \ref{cmd}). Young isolated brown dwarfs are typically slightly redder than older, field objects at the same absolute magnitude \citep{liu2013}. Physically, this redward offset occurs due to the lower surface gravity and higher internal heat of the younger objects. Given the apparent inflation of KELT-1b's atmosphere \citep{siverd2012}, we might expect that it would appear more similar to these younger objects than the higher surface gravity field brown dwarfs 

While our measured $H-K$ color for KELT-1b's day side is consistent with it laying along the general color sequence for brown dwarfs, the uncertainties on  make it difficult to distinguish if the broadband properties of the atmosphere are more similar to a young or a field brown dwarf. We do find it more likely that the day side colors of KELT-1b are that of a field brown dwarf, but this is at a low ($1.6\,\sigma$) significance.

\subsection{The Lack of a Stratopsheric Temperature Inversion}

As mentioned in the introduction, early expectations were that hot Jupiters with day sides hotter than approximately 1900\,K should show a stratospheric temperature inversion in their TP profiles due to the presence of gaseous TiO or VO \citep{fortney2008}. However, subsequent eclipse observations of several giant planets with day side temperatures of 2000\,K to 2500\,K showed no convincing evidence for an inversion in their day side emission spectra \citep{madhu2014,crossfield2015}. Given the theoretical expectation that TiO/VO should be generically present in the atmospheres of giant planets, \cite{spiegel2009} posited that a vertical ``cold-trap'' might strongly deplete the TiO/VO abundance in the upper atmosphere of hot Jupiters. Given the strong day-night contrast observed on hot Jupiters with day sides hotter than 2000\,K \citep[e.g.,][and references therein]{komacek2016}, \cite{parmentier2013} later suggested that a day-night cold-trap would also be able to deplete the TiO/VO abundance in the upper atmospheres of these hot giant planets.

Both a vertical and day-night cold-trap envision TiO and VO raining-out of a hot Jupiter's upper atmosphere, by condensing from the gas phase and then gravitationally settling into the planetary interior. In the case of a vertical cold trap, the day side TP profile is assumed to cross below the TiO or VO condensation curves at some point relatively high in the atmosphere. Gas-phase TiO/VO molecules that happen to fall below this height will then begin to condense and fall further into the atmosphere. Without some vigorous mechanism to loft the TiO/VO condensates back into regions hot enough for TiO/VO gas, \cite{spiegel2009} found that this process would rapidly deplete an upper atmosphere's gas-phase TiO/VO.

A day-night cold-trap is broadly similar, but is instead caused by TiO/VO condensing on the cooler night side of a hot Jupiter. Briefly, if TiO/VO molecules can condense and fall far enough while they cross the cold night side of a planet, they will not be able to diffuse back into the upper atmosphere when they revaporize on the day side. \cite{parmentier2013} found that if TiO/VO were to condense into particles larger than a few microns on the night sides of hot Jupiters, then a day-night cold-trap would efficiently deplete TiO/VO gas in the upper atmosphere.

Interestingly, recent observations of giant planets with day sides closer to 3000\,K using Wide Field Camera 3 (WFC3) on the Hubble Space Telescope are best fit by inverted or isothermal TP profiles. These planets are WASP-12b \citep[2930\,K,][]{stevenson2014b}, WASP-33b \citep[3300\,K,][]{haynes2015} and WASP-103b \citep[2900\,K][]{cartier2017}. In discussing the emission spectrum of WASP-33b, \cite{haynes2015} suggested that it may be that only the very hot giant planets, with day side brightness temperatures near 3000\,K, avoid the cold-trap processes and posses a stratospheric temperature inversion. 

Since our observations strongly indicate that no inversion (and presumably TiO or VO) is present in the upper atmosphere of KELT-1b, it provides a interesting contrast to the observations of the three hot Jupiters. While all four objects (KELT-1b, WASP-12b, WASP-33b, WASP-103b) show roughly the same day side brightness temperatures, KELT-1b possess a surface gravity approximately 20 times greater than the giant planets. If we accept \cite{haynes2015}'s supposition that inversions may only be present on very high temperature planets, KELT-1b's atmosphere indicates that there may also be a surface gravity dependence.

In both the vertical and day-night cold-traps, the condensate particles gravitationally settle into the planetary interior. For simplicity, let us assume that the region of an atmosphere where the condensates begin to form has a low Knudsen number, so that the terminal velocity of the condensate particles is equal to the Stokes velocity
\begin{equation}\label{eq:4110}
V_{\mathrm{term}}=\frac{2a^2g(\rho_\mathrm{p}-\rho)}{9\eta},
\end{equation}
where $a$ is the particle radius, $g$ is the gravitational acceleration, $\rho_\mathrm{p}$ is the particle density, $\rho$ is the atmospheric density, and $\eta$ is the viscosity of the gas. Let us also assume that the atmospheric compositions of hot giant planets and brown dwarfs are roughly similar, such that the atmospheric scale height on both the day and night side of a hot Jupiter increases with increasing day side temperature, and that the night side crossing and the vertical mixing time are roughly independent of both temperature and surface gravity in the regimes we are interested in. Then the time for a particle to free fall one atmospheric scale height, $H=k_{\mathrm{B}}T/\mu_{\mathrm{m}}g$, will be proportional to
\begin{equation}\label{eq:4120}
\tau_\mathrm{ff}=\frac{H}{V_{\mathrm{term}}} \propto Ta^{-2}g^{-2}.
\end{equation}
Note that since the typical hydrodynamical relaxation time for a giant planet ($\sim$20 minutes) is much shorter than the expected hemisphere crossing time ($\sim$10s hours), the scale height of the atmosphere, and hence $T$ above, on the day and night side will be different.

Under these assumptions, the free-fall timescale will scale linearly with temperature and inversely as the square of the surface gravity. Since both cold-trap processes rely upon condensates free-falling out of the upper atmosphere, to first order the efficiency of the cold-trap processes should go as the inverse of the free-fall time. This possible functional dependence would explain why only hot Jupiters with day side temperatures near 3000\,K have shown evidence for inversions, while KELT-1b -- with its much higher gravity -- does not. Indeed, if free-fall driven cold-traps are the primary inhibitor of inversions within hot giant planet atmospheres, we would not expect to see an inversion in a hot brown dwarf atmosphere unless it was extremely (perhaps prohibitively) hot. On the other hand, stratospheric inversions could be possible in giant planets with day sides closer to 2000\,K if their surface gravities were low.  

Based on our observations of KELT-1b's atmosphere, surface gravity appears to play a noticeable role in setting the observable properties of a hot Jupiter's atmosphere. While we have focused on the presence of a stratospheric temperature inversion in the upper atmosphere, we note that \cite{stevenson2016} recently showed that the cloudiness of hot Jupiters near their terminators appears to depend upon both temperature and surface gravity. \cite{stevenson2016} found that hot, high gravity giant planets show little to no evidence for clouds in transmission spectroscopy, which he suggests is a result of a varying efficiency of vertical diffusion. If this is correct, the different functional dependence that \cite{stevenson2016} finds for clouds compared to our Equation \ref{eq:4120} suggests that the mechanisms underlying cloud formation and stratospheric inversion in hot Jupiter atmospheres -- while both possibly dependent upon surface gravity -- could be fundamentally different.  

\section{Summary}

Using the LUCI1 NIR spectrograph on LBT, we spectrophotometrically observed a single secondary eclipse of KELT-1b on the night of UT 2013 October 26 in the $H$-band. After subtracting the telluric background emission and accounting slight changes in the wavelength solution over the course of the observations, we extracted a set of broadband and spectrally-resolved light curves.

We fit an eclipse model to the broadband data using a Gaussian Process regression analysis, which gave us a clear measurement of the eclipse at $\Delta H=1418\pm94$\,ppm. For the spectrally-resolved data, we divided the time series in each spectral wavechannel by the broadband time series, and measured a set of spectrally-resolved eclipse depths. Using a similar Gaussian Process regression model to the broadband analysis, we are able to divide the $H$-band into five spectral channels and measure the eclipse depth in each to an average precision of approximately $135$\,ppm (Table \ref{tab:specresults}). We find that the average brightness temperature of the day side is approximately $3420\pm70$\,K in the $H$-band, which is hotter than the $3220\pm50$\,K average brightness temperature calculated using all the measured eclipses.

Based on these measurements, and previous broadband eclipse observations at other wavelengths, we find that the day side atmosphere of KELT-1b appears identical to an isolated M5 star or brown dwarf (Figure \ref{emission}). Our modeling of KELT-1b's atmosphere using hot Jupiter atmospheric models yields a qualitatively similar result (Figure \ref{modelatmo}). We therefore conclude that the day side of KELT-1b possesses a monotonically decreasing temperature-pressure profile, and is neither isothermal, nor does it posses a stratospheric temperature inversion.

This is in contrast to the hot Jupiters WASP-12b \citep{stevenson2014b}, WASP-33b \citep{haynes2015} and WASP-103b \citep{cartier2017}, all three of which have similar dayside brightness temperatures as KELT-1b, but all show evidence for either an isothermal or inverted atmospheric TP profile. Our observations of KELT-1b indicate an additional surface gravity dependence on the processes governing the presence of an inversion.

\cite{spiegel2009} and \cite{parmentier2013} have previously suggested that vertical or day-night cold-traps in hot Jupiters' atmospheres may be responsible for depleting the abundance of TiO and VO in these planets' upper atmospheres, inhibiting inversions. To first order, the efficiency of a cold-trap should scale as the inverse of this free-fall timescale, and based on this we hypothesize that  KELT-1b's very high surface gravity dramatically shortens the free-fall timescale, thereby allowing TiO/VO to be ``cold-trapped'' out of the atmosphere.

Our observations of KELT-1b are not sufficient to distinguish between a field brown dwarf spectral template with high surface gravity, and a low gravity spectral template. On an $H-K$ color-magnitude diagram, the day side of KELT-1b appears to lie directly on the field brown dwarf color sequence. However, the uncertainties on KELT-1b's $H$ and $K$ eclipse depths mean that our measurement of the day side's $H-K$ color is only $1.6\,\sigma$ off from the low-gravity color sequence.

In the coming year, observations of KELT-1b's eclipse using Wide Field Camera 3 on the Hubble Space Telescope should allow us to differentiate between the field and low-gravity spectral templates. Since KELT-1b transits its host star, we are able to independently measure its mass and radius, and thus its surface gravity, without relying on template matching or gravity indices. These upcoming HST/WFC3 will therefore, hopefully, be the first direct test of the spectral surface gravity indicators used throughout the fields of brown dwarfs and directly-imaged planets. We are also in the process of analyzing Spitzer phase curve observations of KELT-1b at \three and \four, which should allow us to see how the atmosphere changes from the day to night side. It will be interesting to see if the striking similarity of the day side to isolated field brown dwarfs also hold true on the night side of KELT-1b.

\acknowledgements
We thank the anonymous referee for their comments during the review process, and we wish to thank Leo Liu, Jason Wright, and Ming Zhao for helpful discussions. 

T.G.B. partially supported by funding from the Center for Exoplanets and Habitable Worlds. The Center for Exoplanets and Habitable Worlds is supported by the Pennsylvania State University, the Eberly College of Science, and the Pennsylvania Space Grant Consortium. Work by T.G.B and B.S.G. was partially supported by NSF CAREER Grant AST-1056524. 

The LBT is an international collaboration among institutions in the United States, Italy and Germany. LBT Corporation partners are: The Ohio State University, The University of Arizona on behalf of the Arizona university system; Istituto Nazionale di Astrofisica, Italy; LBT Beteiligungsgesellschaft, Germany, representing the Max-Planck Society, the Astrophysical Institute Potsdam, and Heidelberg University; The Research Corporation, on behalf of The University of Notre Dame, University of Minnesota and University of Virginia.

This publication makes use of data products from the Two Micron All Sky Survey, which is a joint project of the University of Massachusetts and the Infrared Processing and Analysis Center/California Institute of Technology, funded by the National Aeronautics and Space Administration and the National Science Foundation.

This work has made use of NASA's Astrophysics Data System, the Extrasolar Planet Encyclopedia at exoplanet.eu \citep{exoplanetseu}, the SIMBAD database operated at CDS, Strasbourg, France,  and the VizieR catalog access tool, CDS, Strasbourg, France \citep{vizier}.

\begin{deluxetable*}{lcl}
\tablecaption{Prior Values for the Broadband Eclipse Parameters}
\tablehead{\colhead{~~~Parameter} & \colhead{Units} & \colhead{Value}}
\startdata
              $T_S$\dotfill &Predicted Eclipse Time (\bjdtdb)\dotfill & $2456591.69888\pm0.00075$\\
                 $\log(P)$\dotfill &Log orbital period (days)\dotfill & $0.0854668\pm0.0000014$\\
                            $\sqrt{e}\cos{\omega}$\dotfill & \dotfill & $0.005\pm0.03$\\
                            $\sqrt{e}\sin{\omega}$\dotfill & \dotfill & $0.001\pm0.075$\\
                     $\cos{i}$\dotfill &Cosine of inclination\dotfill & $0.052\pm0.023$\\
     $R_{P}/R_{*}$\dotfill &Radius of planet in stellar radii\dotfill & $0.077590\pm0.00058$\\
$\log(a/R_{*})$\dotfill &Log semi-major axis in stellar radii\dotfill & $0.559\pm0.006$
\enddata
\tablecomments{These values are the average of the parameters from the dual-channel \emph{Spitzer} eclipse measurements in \cite{beatty2014}, except for the period measurement, which we updated using transit measurements described in Section 3.2.}
\label{tab:broadbandpriors}
\end{deluxetable*}

\begin{deluxetable*}{lcl}
\tablecaption{Median Values and 68\% Confidence Intervals for the Broadband Eclipse}
\tablehead{\colhead{~~~Parameter} & \colhead{Units} & \colhead{Value}}
\startdata
\sidehead{GP Hyperparameters:}
             	          $A_t$\dotfill &Covariance Amplitude\dotfill & $0.0012_{-0.0011}^{+0.0397}$\\
                       $L_t$\dotfill &Covariance Length-scale\dotfill & $0.257_{-0.123}^{+0.376}$\\
\sidehead{Eclipse Model Parameters:}
                        $T_S$\dotfill &Eclipse Time (\bjdtdb)\dotfill & $2456591.6958\pm0.0005$\\
                 $\log(P)$\dotfill &Log orbital period (days)\dotfill & $0.0854668\pm0.0000014$\\
                            $\sqrt{e}\cos{\omega}$\dotfill & \dotfill & $0.005\pm0.03$\\
                            $\sqrt{e}\sin{\omega}$\dotfill & \dotfill & $0.004\pm0.073$\\
                     $\cos{i}$\dotfill &Cosine of inclination\dotfill & $0.064\pm0.019$\\
     $R_{P}/R_{*}$\dotfill &Radius of planet in stellar radii\dotfill & $0.07742\pm0.00056$\\
$\log(a/R_{*})$\dotfill &Log semi-major axis in stellar radii\dotfill & $0.557\pm0.005$\\
                        $\delta$\dotfill &Eclipse depth (ppm)\dotfill & $1418\pm94$\\
\sidehead{Derived Parameters:}
                           $P$\dotfill &Orbital period (days)\dotfill & $1.217494\pm0.000004$\\
          $a/R_{*}$\dotfill &Semi-major axis in stellar radii\dotfill & $3.60\pm0.04$\\
                           $i$\dotfill &Inclination (degrees)\dotfill & $86.3\pm1.1$\\
                                $b$\dotfill &Impact Parameter\dotfill & $0.232_{-0.072}^{+0.063}$\\
                     $T_{FWHM}$\dotfill &FWHM duration (days)\dotfill & $0.1061_{-0.0019}^{+0.0014}$\\
               $\tau$\dotfill &Ingress/egress duration (days)\dotfill & $0.00894\pm0.00024$\\
                      $T_{14}$\dotfill &Total duration (days)\dotfill & $0.1151\pm0.0016$\\
                                   $e\cos{\omega}$\dotfill & \dotfill & $0.0002\pm0.002$\\
                                   $e\sin{\omega}$\dotfill & \dotfill & $0.000006\pm0.0058$\\
                            $e$\dotfill &Orbital Eccentricity\dotfill & $0.0037_{-0.0028}^{+0.0080}$\\
           $\omega$\dotfill &Argument of periastron (degrees)\dotfill & $3_{-39}^{+32}$
\enddata
\label{tab:broadbandresults}
\end{deluxetable*}

\begin{deluxetable*}{lccccc}
\tablecaption{Median Values and 68\% Confidence Intervals for the Spectrally-Resolved Eclipses}
\tablehead{\colhead{Parameter} & \colhead{$1.55\mu$m--$1.59\mu$m} & \colhead{$1.59\mu$m--$1.63\mu$m} & \colhead{$1.63\mu$m--$1.67\mu$m} & \colhead{$1.67\mu$m--$1.71\mu$m} & \colhead{$1.71\mu$m--$1.75\mu$m}}
\startdata
\sidehead{Spectrally-Resolved Eclipse Model Parameters:}
$\delta$ (ppm)\dotfill & $1285\pm131$ & $1461\pm128$ & $1587\pm130$ & $1532\pm137$ & $1623\pm158$\\
$T_S$ (\bjdtdb$-2456590$)\dotfill & $1.6957\pm0.0005$ & $1.6959\pm0.0005$ & $1.6957\pm0.0005$ & $1.6958\pm0.0005$ & $1.6959\pm0.0005$\\
\sidehead{GP Hyperparameters:}
$A_\mathrm{total}$ ($\times10^{-5})$\dotfill & $1.6\pm1.5$ & $5.9\pm3.7$ & $7.8\pm4.3$ & $1.5\pm1.2$ & $2.7\pm1.5$\\
$L_t$\dotfill & $0.23\pm0.02$ & $0.19\pm0.03$ & $0.18\pm0.025$ & $0.15\pm0.03$ & $0.429\pm0.05$\\    
$L_\mathrm{air}$\dotfill & $2.06\pm0.59$ & $1.39\pm0.35$ & $1.48\pm0.23$ & $1.78\pm0.85$ & $1.56\pm1.05$                   
\enddata
\label{tab:specresults}
\end{deluxetable*}

\end{document}